\documentstyle[12pt]{article}
\newlength{\extraspace}
\newlength{\extraspaces}
\newcommand{\be}{\begin{equation}
\addtolength{\abovedisplayskip}{\extraspaces}
\addtolength{\belowdisplayskip}{\extraspaces}
\addtolength{\abovedisplayshortskip}{\extraspace}
\addtolength{\belowdisplayshortskip}{\extraspace}}
\newcommand{\ee}{\end{equation}}

\newcommand{\ba}{\begin{eqnarray}
\addtolength{\abovedisplayskip}{\extraspaces}
\addtolength{\belowdisplayskip}{\extraspaces}
\addtolength{\abovedisplayshortskip}{\extraspace}
\addtolength{\belowdisplayshortskip}{\extraspace}}
\newcommand{\ea}{\end{eqnarray}}

\newcommand{\newsection}[1]{
\vspace{15mm}
\pagebreak[3]
\addtocounter{section}{1}
\setcounter{equation}{0}
\setcounter{subsection}{0}
\setcounter{footnote}{0}
\begin{flushleft}
{\large\bf \thesection. #1}
\end{flushleft}
\nopagebreak
\medskip
\nopagebreak}

\newcommand{\newsubsection}[1]{
\vspace{1cm}
\pagebreak[3]

\addtocounter{subsection}{1}
\noindent{ \bf \thesubsection. #1}
\nopagebreak
\vspace{2mm}
\nopagebreak}

\newcommand{\ie}{{\it i.e.\ }}

\newcommand{\hf}{{\textstyle{1\over 2}}}
\newcommand{\tr}{{\rm tr}}
\newcommand{\is}{\! & \! = \! & \!}

\newcommand{\nonu}{\nonumber \\[1.5mm]}

\newcommand{\cd}{\! \cdot \!}

\newcommand{\half}{\textstyle{1\over 2}}

\newcommand{\ra}{\rightarrow}

\newcommand{\ab}{{\alpha\beta}}

\newcommand{\mint}{\! - \!}

\newcommand{\fijn}[1]{
\begin{figure}
\begin{center}
\setlength{\unitlength}{.75mm}
\raisebox{-40\unitlength}
{\mbox{\begin{picture}(80,23)(-10,-17)
\thicklines
\put(-45,-10){\line(0,1){20}}
\put(-45,1){\vector(0,-1){2}}
\put(-47,8){\makebox(0,0){\footnotesize{$\theta$}}}
\put(-47,-8){\makebox(0,0){\footnotesize{$\phi$}}}
\put(-10,-10){\line(0,1){20}}
\put(-10,1){\vector(0,-1){2}}
\put(0,-10){\line(0,1){20}}
\put(-0,1){\vector(0,-1){2}}
\put(35,-10){\line(0,1){20}}
\put(35,1){\vector(0,-1){2}}
\put(45,-10){\line(0,1){20}}
\put(45,1){\vector(0,-1){2}}
\put(55,-10){\line(0,1){20}}
\put(55,1){\vector(0,-1){2}}
\put(85,-10){\line(0,1){20}}
\put(85,7){\vector(0,-1){2}}
\put(85,-4){\vector(0,-1){2}}
\put(100,-10){\line(0,1){20}}
\put(100,7){\vector(0,-1){2}}
\put(100,-4){\vector(0,-1){2}}
\thinlines
\put(80,10.5){\line(1,0){25}}
\put(80,-10.5){\line(1,0){25}}
\put(80,10){\line(1,0){25}}
\put(80,-10){\line(1,0){25}}
\put(85,0){\line(1,0){15}}
\put(-50,10){\line(1,0){10}}
\put(-50,-10){\line(1,0){10}}
\put(-15,10){\line(1,0){20}}
\put(-15,-10){\line(1,0){20}}
\put(30,10){\line(1,0){30}}
\put(30,-10){\line(1,0){30}}
\put(-50,10.5){\line(1,0){10}}
\put(-50,-10.5){\line(1,0){10}}
\put(-15,10.5){\line(1,0){20}}
\put(-15,-10.5){\line(1,0){20}}
\put(30,10.5){\line(1,0){30}}
\put(30,-10.5){\line(1,0){30}}
\end{picture}}}
\parbox{13cm}{\small #1}
\end{center}
\end{figure}}

\newcommand{\asympt}[1]{
\begin{figure}
\begin{center}
\setlength{\unitlength}{.95mm}
\raisebox{-40\unitlength}
{\mbox{\begin{picture}(80,66)(-40,-40)
\thicklines
\put(-30,0){\line(1,1){30}}
\put(0,30){\line(1,-1){30}}
\put(30,0){\line(-1,-1){30}}
\put(0,-30){\line(-1,1){30}}
\thinlines
\put(0,18){\vector(1,0){9}}
\put(0,18){\vector(-1,0){9}}
\put(0,-18){\vector(1,0){9}}
\put(0,-18){\vector(-1,0){9}}
\put(18,0){\vector(0,1){9}}
\put(18,0){\vector(0,-1){9}}
\put(-18,0){\vector(0,1){9}}
\put(-18,0){\vector(0,-1){9}}
\put(-18,-17){\makebox(0,0){$h_1$}}
\put(-18,17){\makebox(0,0){${g_2}$}}
\put(19,-17){\makebox(0,0){${g_1}$}}
\put(20,17){\makebox(0,0){${h_2}$}}
\put(12,0){\makebox(0,0){\footnotesize{$\log s$}}}
\put(0,14){\makebox(0,0){\footnotesize{$\log (se^{i\pi})$}}}
\put(-12,0){\makebox(0,0){\footnotesize{$\log s$}}}
\put(0,-14){\makebox(0,0){\footnotesize{$\log (s e^{i\pi})$}}}
\end{picture}}}
\parbox{13cm}{\small #1}
\end{center}
\end{figure}}
\newcommand{\xper}{z}
\newcommand{\xpar}{x}
\newcommand{\zper}{z_{{}_{\!R}}}
\newcommand{\zpel}{z_{{}_{\!L}}}
\newcommand{\del}{\partial}
\newcommand{\ppar}{{{}_{\! /\! /}}}
\newcommand{\pper}{{{}_{\! \perp}}}

\newcommand{\twop}[1]{
\begin{figure}
\begin{center}
\setlength{\unitlength}{1.00mm}
\raisebox{-40\unitlength}
{\mbox{\begin{picture}(80,45)(-35,-30)
\thicklines
\put(-22,22){\line(1,-1){41}}
\put(-15,15){\vector(-1,1){1}}
\put(16,23){\line(-1,-1){42}}
\put(9,16){\vector(1,1){1}}
\put(0,0){\vector(-2,-1){13}}
\thinlines
\put(-8,0){\line(1,0){35}}
\put(8,4){\line(-2,-1){35}}
\put(0,-8){\line(0,1){35}}
\put(27,0){\vector(1,0){1}}
\put(0,27){\vector(0,1){1}}
\put(-13,20){\makebox(0,0){$V^-$}}
\put(18,20){\makebox(0,0){$V^+$}}
\put(-6,-7){\makebox(0,0){$z$}}
\put(25,-2){\makebox(0,0){$x$}}
\put(-2,25){\makebox(0,0){$t$}}
\end{picture}}}
\parbox{13cm}{\small #1}
\end{center}
\end{figure}}

\newcommand{\xe}{x_{{}_{\! 1}}}
\newcommand{\xt}{x_{{}_{\! 2}}}
\newcommand{\xet}{x_{{}_{\! 12}}}

\newcommand{\ze}{z_{{}_{\! 1}}}
\newcommand{\zt}{z_{{}_{\! 2}}}
\newcommand{\zet}{z_{{}_{\! 12}}}

\begin{document}

\addtolength{\baselineskip}{.8mm}

\thispagestyle{empty}

\begin{flushright}
{\sc PUPT}-1319\\
Sept 93
\end{flushright}
\vspace{.3cm}

\begin{center}
{\large\sc{QCD at High Energies\\[6mm]
and Two-Dimensional Field Theory}}\\[7mm]

{\sc  Herman Verlinde}\\[3mm]
{\it Joseph Henry Laboratories\\[2mm]
Princeton University, Princeton, NJ 08544} \\[.3cm]
{ and}\\[.3cm]

{\sc Erik Verlinde}\\[3mm]{\it TH-Division, CERN\\[2mm]
CH-1211 Geneva 23, Switzerland\\[2mm] and\\[2mm] Institute for
Theoretical Physics,
Utrecht University,\\[2mm]  P.O.Box 80.006,  3508 TA Utrecht\\[2mm]
The Netherlands}\\[7mm]

{\sc Abstract}
\end{center}

\noindent
Previous studies of high-energy scattering in QCD have shown a
remarkable correspondence with two-dimensional field theory.
In this paper we formulate a simple effective model in which this
two-dimensional nature of the interactions is manifest. Starting from
the (3+1)-dimensional Yang-Mills action, we implement the high energy
limit $s\! >\! > \! t$ via a scaling argument and we derive from this
a simplified effective theory. This effective theory is still
(3+1)-dimensional, but we show that its interactions can to leading
order be summarized in terms of a two-dimensional sigma-model defined on the
transverse plane. Finally, we verify that our formulation is consistent
with known perturbative results.

\vfill

\newpage
\newsection{Introduction.}

\noindent
High-energy scattering in quantum chromodynamics (QCD) has been the
subject of intensive study since the early seventies \cite{lipetal,chengwu}.
In particular within the framework of perturbation theory, systematic
procedures have been developed for extracting the large $s$, fixed $t$,
behaviour of each amplitude and for summing these contributions using
a leading-log or eikonal approximation scheme \cite{lipetal,chengwu}.
A striking feature of the results obtained by these methods is that the
contributions at each order
take the form of two-dimensional amplitudes, and it has indeed
been suspected for some time that there exists
an intimate relationship between QCD at high
energies and two-dimensional field theory \cite{liplast}.

The two-dimensional nature of the interaction can be understood
semiclassically as follows. Introduce two light-cone
coordinates and two transverse coordinates
\ba
\xpar^\alpha \is (x^+,x^-) \nonu
\xper^i \is (y, z)
\ea
with $x^\pm = x \pm t$, and let us assume that two fast moving
particles  have very
large momenta in the $x^\pm$ direction, while they
remain at a relatively large distance in the $z$-direction.
Now if we consider the  (color) electric field of one of these
fast moving charges, it is clear that due to the
Lorentz-contraction it will take
the form of a shockwave: the field-strength will vanish
everywhere, except on a null-hyperplane through
the trajectory of the particle. The only physical effect the
shockwave can have is that, when a charged test particle passes through it,
its wave function $\psi$ will undergo some instantaneous gauge rotation
$\psi \ra g(z) \psi$. This gauge rotation only
depends on the transverse distance between the particles, and thus
it indeed appears that all interactions essentially take place within
the transverse plane of the shockwave.\footnote{For the case the particles
have only electro-magnetic interactions,
this shockwave interaction has been used in \cite{thooft,jackiwetal}
to give an elegant derivation of the high energy scattering amplitude,
reproducing the result obtained via the eikonal approximation
\cite{eikonal,kabatetal}. Related work in gravity has been done in
\cite{thooft,grav,us}.}

Motivated by this simple physical picture, we will in this paper
propose a new formulation of high-energy elastic scattering
in gauge theory, in which the two-dimensional nature of the interactions
is manifest. The method will have some similarities with the eikonal
approximation, but appears to be more general. We will interested in the
kinematical regime where $s$ is much larger than $t$, while
$t$ is also larger that the QCD scale $\Lambda_{qcd}$. In this case
it is a good approximation to represent the scattering process
between two hadrons as a collection of separate collisions between the
individual quarks, and we will concentrate on these individual quark-quark
scattering processes. The main simplifying assumption we will make is
that all these individual collisions will take place at a very small time
and longitudinal length scale of the order of $1/\sqrt{s}$.
Of course, in actual events there will also be very non-trivial dynamics
taking place at larger scales, since a scattered quark will interact
via secondary collisions with the other quarks inside the hadron, which
as a result will fall apart into one or more jets.
However, it seems reasonable to assume that the amplitude for the
primary collision between the two energetic quarks is independent
of the precise details of these subsequent processes, precisely because
these occur at a much larger time and length scale.

In section 2 we will use this assumption to derive a simplified model
for QCD at high energies. Via a simple scaling argument, similar to the
one used in \cite{us}, we will isolate from the Yang-Mills action the part
of the theory that appears relevant for the dynamics in this regime.
Next, in section 3 we will use the reduced theory to reformulate the
calculation of elastic scattering amplitudes in terms of a two-dimensional
effective field theory, which has the form of a sigma-model defined on
the transverse plane. We make contact with some of the results obtained by
more standard methods \cite{lipetal,chengwu} in a concluding section.
In an appendix we describe the form of the
non-abelian shock-wave solutions, that provide the semiclassical
interpretation of the scattering process.

\newsection{QCD in the high-energy limit.}

\noindent
In this section we will formulate a simple effective description of
QCD at high energies.  We will start from the hypothesis that
the typical longitudinal
momentum of the dynamical modes in this process grows proportional
to the center of mass energy, whereas the typical size of the
transversal momenta is determined by the momentum transfer. The same
assumption also underlies other approaches to high energy QCD, where
it is in particular used to simplify the evaluation of Feynman diagrams.
Here, however, we wish to take this high energy limit directly at the
level of the action. This procedure has the important advantage
that one does not need to fix the gauge before taking the
limit, and thus the final effective theory will still be gauge invariant.

\newsubsection{The scaling argument.}

\noindent
We first consider the pure Yang-Mills model, described by
\be
\label{ym}
S =  {\textstyle {1\over 4}} \int\!d^4x\,
\tr (F_{\mu\nu}F^{\mu\nu}).
\ee
Here $F_{\mu\nu}$ is the non-abelian field strength
\be
\qquad \qquad F_{\mu\nu} =
\partial_\mu A_\nu \! - \partial_\nu A_\mu + e \, [A_\mu,A_\nu] , \qquad \qquad
A_\mu = A_\mu^a \tau^a
\ee
where $e$ is the coupling constant and
$\tau^a$ are the generators of the lie algebra of the
gauge group $G = SU(N)$.
To extract that part of the action that is
relevant to the high energy forward
scattering of quarks,
let us consider the behaviour of (\ref{ym})
under a rescaling of the longitudinal coordinates
\be
\label{re}
x^\alpha \ra  \lambda x^\alpha,
\ee
The idea is that by performing this rescaling inside the Yang-Mills
action, we will see which part will become strongly or weakly coupled,
when we look at the theory at high longitudinal energies.
The components of the gauge potential transform under (\ref{re})
as $A_i \ra A_i$, while $A_\alpha \ra \lambda^{-1} A_\alpha$.
Hence the rescaled Yang-Mills action can be written in the following form
\be
\label{resca}
S_{YM}^\prime = {1\over 2} \int \tr ( E^{\ab}F_\ab
+F_{\alpha i}F^{\alpha i}) + {\lambda^2\over 4}
\int \tr(E_{\ab}E^{\ab}+F_{ij}F^{ij}),
\ee
Here we introduced $E_{\ab}=-E_{\ab}$ as an auxiliary field.
So far we have done nothing. Indeed, the
description of a scattering process with some $s$ and $t$ using
the standard action is completely equivalent to that using the
rescaled action (\ref{resca}), provided we also rescale $s$ to
$$
s'= \lambda^2 s.
$$
The point, however, is that we can now use this correspondence
and choose
$$
\lambda \sim {1\over \sqrt{s}} \ra 0,
$$
so that $s'$ is kept fixed, and thus reformulate the high energy
limit $s\ra \infty$ in QCD as the $\lambda \ra 0$ limit of the
rescaled theory (\ref{resca}).

The parameter $\lambda$ is not a coupling constant of the original theory,
but introduced via the rescaling, so what does it
mean to take the limit $\lambda \ra 0$?
Here we must go back our basic starting point, namely that, before the
rescaling (\ref{re}), the typical longitudinal momentum of the
dynamical modes is much larger than the typical transversal momentum.
For these modes the second term in (\ref{resca})
is indeed subdominant, and so, the $\lambda \ra 0$ limit in fact
corresponds to a {\it truncation} of the full theory to these high
energy modes. By our assumption, the contribution of the
modes removed by this truncation is subleading for $s \! >\!\! >\! t$.

Further it seems reasonable to assume that in leading order in $\lambda$
we can neglect the second part of the
action (\ref{resca}). However, we need to be a little careful
here, since one could imagine that the $\lambda \ra 0$
limit of the amplitudes of the full theory are not equal to those
of the $\lambda = 0$ theory.
The reduced theory could for example be singular, in which case
the second term in (\ref{resca}) must be kept as a regulator.
Keeping this in mind, let us however for the moment assume
that the $\lambda = 0$ theory is well-defined as it stands, in which
case it must give the leading order contribution.

Thus, we propose that high energy scattering
in gauge theories can be described by means of the following truncated
Yang-Mills action
\be
\label{Sred}
S[A] ={1\over 2}\int 
\tr(E^{\ab}F_\ab
+ F_{\alpha i}F^{\alpha i} ).
\ee
In this theory we wish to calculate the quark-quark scattering amplitude.
The coupling to charged quark fields $\psi$ is described by
the high energy limit of the usual quark action, which takes the form
\be
S[\psi,A] = \int 
\overline{\psi}\gamma^\alpha(\partial_\alpha + e A_\alpha) \psi.
\ee
Again, this action is obtained from the standard action by performing the
rescaling (\ref{re}) and taking limit $\lambda \ra 0$. Hence we see that,
as was to be expected of a
high energy limit, the quarks propagate only in the longitudinal
direction and only couple to the components $A_\alpha$ of the gauge
potential. Note further that the mass term is subleading, so
we can divide $\psi$ into left- and right moving components.
It should again
be mentioned, however, that the subleading terms of the action
could still be important for regulating possible singularities due to
the fact that the leading order model is ultra-local in the transverse
$z$-direction.

\newsubsection{Relation with Lipatov's gluon emission vertex.}

\noindent
The first property of the leading order theory described by (\ref{Sred})
we notice is that the auxiliary field $E^\ab$ has become a Lagrange
multiplier imposing the zero-curvature constraint
\be
F_{+-}=0.
\ee
In other words, in the high energy limit, the leading order contribution
comes from those gauge field configurations that are flat in
the longitudinal direction. This is the central observation of this paper.
As the above reasoning shows, it is a simple and direct consequence of the
fact that we are interested in processes for which the typical longitudinal
scale is much smaller than the typical transversal scale.

\newcommand{\lip}[1]{
\begin{figure}
\begin{center}
\setlength{\unitlength}{1mm}
\raisebox{-40\unitlength}
{\mbox{\begin{picture}(50,23)(-10,-17)
\thicklines
\put(10,12.25){\vector(1,0){2}}
\put(10,-12.25){\vector(1,0){2}}
\put(10,0){\vector(1,0){2}}
\put(5,8){\vector(0,-1){2}}
\put(5,-5){\vector(0,-1){2}}
\thinlines
\put(5,-12){\line(0,1){24}}
\put(2,7){\makebox(0,0){$k_1$}}
\put(2,-6){\makebox(0,0){$k_2$}}
\put(10,15){\makebox(0,0){$p_+$}}
\put(10,-15){\makebox(0,0){$p_-$}}
\put(0,12.5){\line(1,0){22}}
\put(0,-12.5){\line(1,0){22}}
\put(0,12){\line(1,0){22}}
\put(0,-12){\line(1,0){22}}
\put(5,0){\line(1,0){15}}
\end{picture}}}
\parbox{13cm}{\small #1}
\end{center}
\end{figure}}

At first sight this observation seems to be in contradiction with
the effective high-energy gluon emission vertex derived by Lipatov,
which produces gluon-radiation with $F_{+-}\neq 0$.
To clarify this point we will briefly summarize the derivation of
Lipatov's vertex, first from the conventional Yang-Mills action, and
then from the high energy effective theory. Lipatov's vertex describes
the effective gluon production relevant to high energy scattering and is
given by \cite{lipatov} 
\ba
\label{lipvert}
C_i(k_1,k_2) \is -(k_1+k_2)_i \nonu
C_+(k_1,k_2) \is (\alpha + 2 {k_{1}^2\over \beta s})p_+ \\[.5mm]
C_-(k_1,k_2) \is -(\beta + 2 {k_{2}^2 \over \alpha s})p_- \nonumber
\ea
Here $p_+$ and $p_-$ denote the momenta of the left and right-moving
quarks, $s = 2p_+p_-$, and $k_1 = k_{1,\pper} + \alpha p_+$ and
$k_2 = k_{2,\pper} + \beta p_-$ are the respective
momenta of the intermediate gluons (see fig 1.)
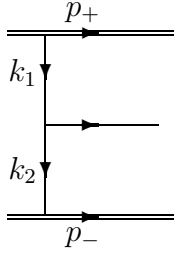
\lip{Fig 1. This figure explains the notation used in the
expression (\ref{lipvert}) of the effective gluon production vertex.
The double horizontal lines represent the fast quarks and the
other are gluon lines.}

To derive this result, let us consider the Yang-Mills equations in the
presence of a source $j_\alpha$, that represents the energetic
quarks. We will implement the high energy limit by
assuming that it is chirally conserved
\be
D_+j_- = D_- j_+ = 0
\ee
Next we solve the Y-M equations perturbatively by expanding
the gauge potential $A_\mu$ and the source $j_\alpha$ in powers of
the coupling. For our purpose we will need to go to the next to
leading order. Given the leading order solutions
\ba
A_i^{(0)} \is 0 \nonu
\label{lo}
\del^2_\pper A_\alpha^{(0)} \is j_\alpha^{(0)}
\ea
the next to leading order equations become
(in the Lorentz gauge)
\ba
\label{avoor}
\del^2  A^{(1)}_i \is [ A^{(0)}_\alpha,
\del_i A^{(0)}_\alpha ] \\[.5mm]
\label{ana}
\del^2 A^{(1)}_\alpha
\is  j_\alpha^{(1)} + \lambda^{-2}
[ A_\beta^{(0)},\del_\beta A_\alpha^{(0)}]
\ea
with
$\del^2 =  \lambda^{-2}\del_\ppar^2 + \del_\pper^2$ and where
in the second equation we used that
$\epsilon^{\ab} \partial_\alpha A^{(0)}_\beta = 0$.
The next to leading order contribution $j^{(1)}_\alpha$ to the current
in (\ref{ana}) can be eliminated in terms
of $A_\alpha^{(0)}$, via the conservation equations
\ba
\label{ja}
\partial_- j^{(1)}_+ \is [A^{(0)}_-, j^{(0)}_+] \nonu
\partial_+ j^{(1)}_- \is [A^{(0)}_+, j^{(0)}_-]
\ea
Using (\ref{lo}) this leads to
\ba
j^{(1)}_+ \is [ {1\over \del_-} A_-^{(0)},  \del_\pper^2  A^{(0)}_+] \nonu
j^{(1)}_- \is [ {1\over \del_+} A_+^{(0)},  \del_\pper^2  A^{(0)}_-]
\ea
which, inserted into (\ref{ana}), gives
\ba
\label{lipvertex}
\del^2 A^{(1)}_+
\is [ {1\over \del_-} A_-^{(0)},  \del_\pper^2  A^{(0)}_+] +
 \lambda^{-2}[ A_-^{(0)},\del_+ A_+^{(0)}]\nonu
\del^2 A^{(1)}_-
\is  [ {1\over \del_+} A_+^{(0)},  \del_\pper^2  A^{(0)}_-] +
 \lambda^{-2}[ A_+^{(0)},\del_- A_-^{(0)}]
\ea
Combined with (\ref{avoor}) this is, when translated to momentum space,
identical to the effective high energy
gluon emission vertex of Lipatov.

Note that if we compute the longitudinal part of the Yang-Mills curvature
$F_{+-}$, to this order, we find a non-zero result
\be
\del_+ A^{(1)}_--\del_- A^{(1)}_+ -  [A^{(0)}_+, A^{(0)}_-]
= 2\lambda^2
{1\over \del_\ppar^2} [\del_i A^{(0)}_+,\del_i A^{(0)}_-]
\ee
As expected, this non-zero result is proportional to (square of) the scaling
parameter $\lambda$, so that it vanishes in the $\lambda \ra 0$ limit.
However, we are not allowed to drop its contribution, because
$F_{+-}$ enters in the equation of motion
with a prefactor of $\lambda^{-2}$.

Now let us consider the analogous computation in the effective theory
with $\lambda = 0$.
To leading order nothing changes, as the solution (\ref{lo})
for $A_\alpha^{(0)}$ automatically satisfies $F_{+-} = 0$.
The next to leading terms in the equation of motion now read
\ba
\del_+ A^{(1)}_--\del_+A^{(1)}_+ \is
[A_+^{(0)},A_-^{(0)}] \nonu
\del_\pper^2 A_\pm^{(1)} \is j_\pm^{(1)} \pm
\partial_\pm E^{(1)}
\ea
Here the first equation is the flatness equation $F_{+-} = 0$ and
the second equation replaces the Yang-Mills equation (\ref{ana}).
Now it may seem that one has made a mistake in putting $\lambda = 0$,
because it implies that $F_{+-} = 0$ exactly, whereas in the full
theory there was a small but non-negligible contribution coming from
the gauge-field variations with $F_{+-}\neq 0$. However, as is usually
the case in these type of situations, the role of the
fluctuations that are transverse to the constraint surface is
taken over by the lagrange multiplier field, in this case $E^{(1)}$.
Indeed, with the help of (\ref{ja}) we can solve for $E^{(1)}$, and find
\be
\del_\ppar^2 E^{(1)} = 2[\del_i A_+,\del_i A_-]
\ee
Hence the value of $E^{(1)}$ precisely equals that of $\lambda^{-2} F_{+-}$
of the finite $\lambda$ theory. It is now not hard to convince oneself that
this correspondence is sufficient to ensure that the perturbation theory
in the $\lambda = 0$ model is equivalent to the high-energy
limit of the convential theory proposed by Lipatov in \cite{liplast},
at least to this order. This correspondence further supports
the validity of our scaling hypothesis.

\newsubsection{Definition of the scattering amplitude.}

Encouraged by this further justification of the high-energy lagrangian
(\ref{Sred}) we will now proceed to study the quark-quark scattering
amplitude in this theory. The calculations simplify drastically due
to the fact that on-shell the gauge-potential $A_\alpha$ is flat.
In particular, it follows that the quark propagator for fixed
gauge potential is simply represented by the Wilson line of
$A_\alpha$. For example,
the two-point function of two right-moving quarks reduces to
\ba
\langle T(\overline{\psi}(\xe,\ze) \psi(\xt,\zt))\rangle_A
\is \delta^{(2)}(\zet)\, \Bigl(\delta(\xet^-)\theta(\xet^+) +
{1 \over \xet^- + i\epsilon} \Bigr) \nonumber \\[2.5mm]
& & \quad \times \, {\rm P}\, \exp(e \, \int_{1}^{2} d x^+ A_+)
\ea
Thus we are led to a description of the scattering amplitude
in which the quarks are represented by light-like Wilson lines,
a prescription that has been proposed previously
by Nachtmann in \cite{nachtmann}. In his derivation the Wilson lines
arise from the coupling of $A_\alpha$ to the quark current by
neglecting the recoil terms in the quark propagators. Our argument
that $A_\alpha$ is predominantly flat gives a new justification
for using this procedure in the high energy limit.

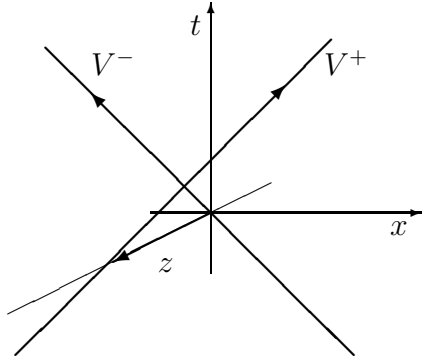
\twop{Fig 2. The configuration of the two light-like
Wilson-line operators representing
the two fast moving quarks.}

A priori, a single
quark-quark scattering-amplitude is by itself not a gauge invariant
quantity, since it depends on gauge-rotations at the end-points of the
Wilson-lines.  One way to get an invariant amplitude would be to
consider the scattering of hadron-like objects such as a quarks connected
by transversal Wilson-lines and integrated against some
hadron wave-function.
However, when $t$ is much larger than $\Lambda_{qcd}$ it appears to be
meaningful to consider the quark-quark scattering processes individually.
In this case the role of the accompanying  quarks and gluons
inside the hadron is to provide a ``frame of reference'' for the
scattered quark with respect to which one can define the amplitude
in an invariant way. To make this concrete, we will follow the procedure
advocated in \cite{nachtmann}, and
use this reference frame to `identify' the two end-points of the
Wilson-line in the sense that gauge-rotations are restricted to
act simultaneously and identically at $+\infty$ and $-\infty$. In this way
by taking the trace of the amplitude in both representations one obtains a
meaningful gauge-invariant quantity, which according to \cite{nachtmann}
corresponds to the diffractive term of the amplitude.

Thus, summarizing, we will define
the high energy quark-quark scattering amplitude by the expectation
value of two light-like Wilson lines
\be
\label{s12} \qquad
\qquad \qquad f(s,t) = {is\over 2m^2}
\int d^2 \! \xper \, \, e^{-iq\cdot \xper}\,
\Bigl\langle V_+(0) \,  V_-(\xper) \Bigr\rangle   \qquad
\qquad t = - q_i^2
\ee
\be
\label{holon}
V_\pm(\xper) = {\rm tr}\Bigl[ {\rm P}\exp(e
\int_{\! - \infty}^{\infty} \!\! dx^\pm\, A_\pm(\xper))\Bigr]
\ee
in the theory described by the high energy limit (\ref{Sred}) of the
QCD lagrangian.
The configuration of the two Wilson line operators is depicted in
fig 2.\footnote{In fact, it turns out to be a
singular limit to take the Wilson lines in (\ref{s12}) exactly
light-like. We will later regularize this `infrared' divergence by allowing
each line to have a small light-like component.}
The kinematical factor ${is\over 2m^2}$ comes from the conventional
normalization of the quark wave-function.


\newsection{Reduction to a Two Dimensional Field Theory.}

\noindent
In this
section we will indicate how this model can be further simplified
and eventually reduced to a two-dimensional effective field theory.
To this end, we will first integrate out the lagrange multiplier
field $E^\ab$ in (\ref{Sred}). This produces a functional constraint
that restricts the integral longitudinal gauge-field components
$A_\alpha$ to potentials of the
form
as
\be
\label{flat}
A_\alpha= {1\over e} \partial_\alpha U U^{-1}
\ee
Here the group element $U$ may still depend on all four coordinates.
It can be seen that
the Jacobian of this replacement of the variables $E^\ab$ and
$A_\alpha$ by $U$ is simply equal to 1. When we substitute
(\ref{flat}) into the action $S$, we get
\be
\label{eff}
\qquad \qquad \qquad S[U,A_i]
= {1\over 2 e^2}\int\, \tr (\partial_\alpha (U^{-1}D_iU))^2
\qquad \qquad
D_i \! = \partial_i \! +\!  e A_i
\ee

So far we
have not chosen any gauge, and the above action is indeed manifestly
gauge invariant. We can
now make use of this fact
to choose a gauge in which the longitudinal
components $A_\alpha$ are completely eliminated via a redefinition of $A_i$ to
$\widetilde{A}_i = {1\over e} U^{-1}D_i U$.
This gauge has one obvious advantage, namely that all {\it local}
interactions in fact disappear!
A subtlety, however, is that, if we wish to be able to send in charged
particles, we should not allow for arbitrary  gauge transformations
at infinity. Some information about the asymptotic values of $U$
will have to survive, since these are the only remaining variables that
couple to the quarks.

Notice that if we had chosen some other more standard gauge,
such as
the Landau gauge $\partial_\mu A^\mu=0$, it would have been rather
difficult to recognize that there are no local interactions. More
close inspection, however, would reveal that in such a gauge the
theory in fact describes a free field, $A_i$, interacting with a
topological field theory, consisting of the group variable $U$ and
the ghost fields. Indeed, it appears that many of the
Feynman diagram computations done in \cite{lipetal} and \cite{chengwu}
can be reinterpreted in this way. We believe that this absence of
local four-dimensional interactions is what underlies the apparent
two-dimensional nature of high energy QCD.

\newcommand{\gl}{g_{{}_{2}}}
\newcommand{\gs}{g_{{}_{1}}}

\newcommand{\hl}{h_{{}_{2}}}
\newcommand{\hs}{h_{{}_{1}}}

\newcommand{\ye}{y_{{}_ {1}}}
\newcommand{\yt}{y_{{}_{2}}}

\renewcommand{\ne}{n_{{}_{1}}}
\newcommand{\nt}{n_{{}_{2}}}

\newcommand{\gA}{g_{{}_{\! A}}}
\newcommand{\hB}{h_{{}_{\! B}}}
\newcommand{\MAB}{M^{{}^{\! AB}}}
\newcommand{\MiAB}{M^{{}^{-1}}_{{}_{AB}}}
\newcommand{\am}{a_i^{{-}}}
\newcommand{\ap}{a_i^{{+}}}
\newcommand{\ha}{\hat{a}}
\newcommand{\hap}{\hat{a}_i^{{+}}}
\newcommand{\eps}{\epsilon}

\newsubsection{The Two-dimensional Model.}

\noindent
We now wish to make this idea more concrete.
Let us denote the values of the field
$U$ at the end-points of the Wilson lines in (\ref{s12})
by $g_1$ and $g_2$ for the right
in- and left out-region, and $h_1$ and $h_2$ for the left in- and
right out-region, resp. (See fig 3.) Since $A_\alpha$ is flat,
the Wilson lines can be expressed in terms of these variables, for example
\be
{\rm P}\exp(e \int_{\! - \infty}^{\infty} \!\! dx^+\, A_+(\xper))=
g_2g_1^{-1}(\xper)
\ee
The group variables $g_A$ and $h_A$ ({\footnotesize{$A$}}=1,2) will not be
held fixed, but will be treated as dynamical fields.
Their physical role is to impose the Gauss-law constraints at the end-points
through their field equations. In addition, as discussed in the Appendix,
these asymptotic modes of $U$ appear naturally in the classical description of
the shockwave configurations associated with the fast-moving quarks.
Our aim is to get a two-dimensional model that describes the dynamics of
these shockwaves.  We further impose that
the transversal gauge-field $A_i$ takes the same value at the two end-point
of each Wilson line, which we will denote be $a^\pm_i(z)$ resp.

It is now clear that  the expectation value (\ref{s12}) of the Wilson lines,
representing the elastic quark-quark scattering amplitude, can in
principle be re-expressed as a correlation function in the two-dimensional
effective theory of the $\xpar$-independent variables $(\gA,\hB,a_i,b_i)$
\be
\label{exp}
\langle V_-(\zpel) V_+(\zper) \rangle
= \int [d\gA  d\hB  da_i^\pm]  \, e^{i S[\gA,\hB,a_i^\pm]} \, \,
\tr\Bigl(\gl \gs^{-1}(\zpel)\Bigr) \, \tr\Bigl(\hl \hs^{-1}(\zper)\Bigr),
\ee
where the two-dimensional effective action $S[\gA,\hB,a_i^\pm]$ is obtained
by integrating out all modes of $A_i$, while keeping its asymptotic boundary
values fixed. This calculation can be performed explicitly, since the integral
over $A_i$ is just gaussian due to the absence of local interactions.
Therefore, in the leading semiclassical approximation the only thing we need to
know is the action of the classical field configuration with the given boundary
conditions. To calculate it we use that
for a solution $A^{cl}_i$ of the classical field equations of (\ref{eff}),
one has.
\be
\label{scl}
S[U,A_i^{cl}] = {1\over 2 e^2} \int \! d^2 z \, \tr\Bigl[
\, \int^{{}^{\infty}}_{{}_{-\infty}} \!\!\!\!\!\! dx^+
\, \partial_+ (U^{-1}D_iU)
\, \times \,
\int^{{}^{\infty}}_{{}_{-\infty}} \!\!\!\!\!\! dx^-\, \partial_-
(U^{-1}D_i U) \Bigr]
\ee
We can now express the right-handside directly in terms of the
asymptotic values for $U$ and $A_i$ as follows
\ba
\label{aspt}
\, \int^{{}^{\infty}}_{{}_{-\infty}}
 \!\!\!\!\!\! dx^-\, \partial_- (U^{-1}D_iU) \is \gl^{-1}D^+_i
\gl - \gs^{-1}D^+_i\gs\nonu
\, \int^{{}^{\infty}}_{{}_{-\infty}} \!\!\!\!\!\!
dx^+\, \partial_+ (U^{-1}D_iU) \is \hl^{-1}D^-_i \hl - \hs^{-1}D^-_i\hs
\ea
where the covariant derivatives $D^\pm_i$ on the
right-hand-side are with respect to $a^\pm_i(\xper)$.
Inserting (\ref{aspt}) into (\ref{scl}) we
find that the effective action of the two-dimensional variables
$(\gA,\hB,a_i^\pm)$ is given by
\be
\label{toedi}
S[\gA,\hB,a_i^\pm] = {1\over 2e^2} \int d^2 \xper \,
\MAB \tr(\gA^{-1}D^+_i \gA   \hB^{-1  }D^-_i \hB )
\ee
where $\MAB$ is the $2\times 2$ matrix
\be
\label{mabo}
\MAB = \left(\begin{array}{cc} 1 & -1 \\
-1 & 1 \end{array}\right)
\ee
and the indices {\small{$A$}} ({\small {$B$}}) $=1,2$ are summed over.

The result (\ref{toedi}) can be derived in an alternative fashion
 by doing the functional integral over gluon field  $A_i$ with
the insertion of delta functionals that keep the value of $A_i$ fixed
at the end-points of the Wilson lines. After representing these
delta-functionals as a Fourier integral
it becomes straightforward to do the gaussian integration.
We leave it to the reader to verify that this gives the same result.
This second method makes clear that the matrix $\MAB$
can be identified with the inverse propagator of the gluon field
between the end-points of the Wilson lines. We will make use of this
remark in the next subsection.

Thus, the matrix $\MAB$ represents
 a discrete version of the wave-operator $\partial_+ \partial_-$, and
thus, in a sense, the two-dimensional model (\ref{toedi}) has
a 3+1-dimensional interpretation in which the longitudinal plane
is replaced by the four end-points of the Wilson lines.
This interpretation also explains why  in the expression (\ref{exp})
there is still a factor of $i$ in front of $S$, even though the transverse
$z$-plane is euclidean. Indeed, without this factor of $i$
the indefinite signature of the action would have
become a problem.

Semi-classically our two-dimensional model can be interpreted as describing
high-energy scattering of quarks via shock-waves.
This interpretation 
becomes clear when we consider the classical
equations of motion of (\ref{toedi})-(\ref{mabo}). In appendix A
we describe the shock wave solutions of the Yang-Mills equations for a single
fast-moving quark. In terms of the above model this corresponds to
the case where we couple one single classical point charge to
$h_2 h_1^{-1}$. It can be easily verified that in this situation
the classical equations of motion of the action
(\ref{toedi}) indeed precisely reduce to (\ref{eom1})-(\ref{kalm}),
provided we use the {\it Ansatz} that $h_2 = h_1$ , $a_i^\pm\!=\! 0$
and $g_1 = 1$. In other words, the
variables $(\gl \gs^{-1})(\zpel)$ and $(\hl \hs^{-1})(\zper)$,
which in our model represent the Wilson lines of
the quarks, are via their own equation of motion equal to
the instantaneous gauge-rotation generated by the shock-wave of the
other quark.

Loop corrections are taken into account by adding to the leading
result the one-loop effective action induced by the quark and
gluon loops. These are given by given by the determinants
\be
\label{logdet}
\log \det(D_\alpha^2 + \lambda^2 \partial_i^2)
\ee
in leading order for $\lambda^2 \ra 0$. These determinants give rise
to the usual 1-loop renormalization of the coupling constant,
as well as to corrections to the leading order
action (\ref{toedi}). It may be possible to use techniques of
two-dimensional conformal field theory to compute these corrections,
in a similar way as was done recently in \cite{nair} in the context
of QCD at high temperatures.

\newsubsection{An infrared problem: How log$\,$s enters.}

\noindent
The above description
of the amplitude appears to be essentially independent of the center of
mass energy  of the two quarks.  This is clearly in
contradiction with the well-known fact that amplitudes in QCD
have a very non-trivial $s$-dependence.  The origin of this dependence
lies in the infrared divergences of 3+1-dimensional gauge theory.
In most standard perturbative treatments of high-energy QCD, these infrared
divergences are taken care of by restricting the rapidities of
the intermediate gluons to lie
in between those of the two fast quarks (see eg.\ \cite{chengwu}).
The size of this rapidity
space grows as $\log s$ and in this way amplitudes acquire
an overall factor proportional to some power of $\log s$, depending on the
number of intermediate gluon propagators.
This infrared problem has a direct counterpart in our two-dimensional
theory (\ref{toedi})-(\ref{mabo}), namely, the matrix $\MAB$ given in
(\ref{mabo}) is not invertible, and as a consequence
some of the fields have a singular propagator.
Hence to make the model non-singular we will need to introduce an infrared
cut-off, in a similar way as in the 3+1-dimensional theory. We can do this
as follows.

\asympt{Fig 3.
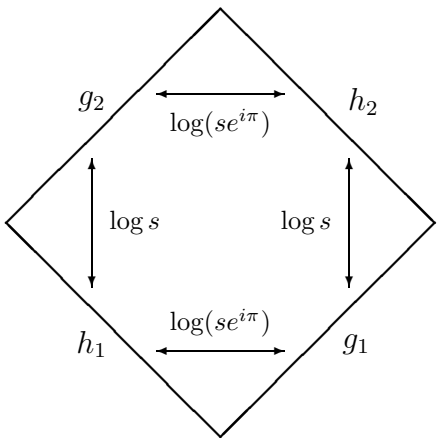
The values of $U$ at the asymptotic regions of the
longitudinal Minkowski plane are denoted by $g_A$ resp. $h_B$,
$A,B \! = \! 1,2$.
The arrows represent the discrete longitudinal propagator
$\MiAB$ given in (\ref{minv}).}

The inverse $\MiAB$ of the matrix $\MAB$ can in fact be identified
with the $A_i$ propagator between the different asymptotic regions of the
longitudinal plane, \ie between the different end-points of the
Wilson-line operators $V_\pm$. The infrared singularity
can be traced back to the fact that the trajectories of
these Wilson-lines were taken to be light-like and therefore have
an infinite distance in rapidity space.
However, for finite quark mass $m$
the classical trajectories of two
quarks with centre of mass energy $s$
are related by a finite Lorentz boost with rapidity parameter $\log {s/m^2}$.
Thus, to regularize this infrared problem, we now give the
Wilson-lines a small timelike component, such that
they coincide with the classical quark trajectories,
and, in addition,  we let them end after some finite proper time $T$.
To be specific, we will choose the end-points of
$V_+$ at $(x_+,x_-)=  \pm\hf T (p_+ , m^2/p_+)$ and
of  $V_-$ at $(x_+,x_-)=  \pm\hf T (m^2/p_-,p_-)$,
with $2p_+ p_- = s$. Note that in the centre of mass frame
$p_+=p_- =\sqrt{s/2}$ so that in the limit $s \ra \infty$,
the trajectories of $V_\pm$ indeed become infinitely long
and light-like.

The next step is to evaluate the gluon
propagator between these end-points.
{}From the fact that the 
propagator $\langle T(A_i(x,z) A_j(0,0)) \rangle$
is proportional to $\log(x^+ x^- \! + i\epsilon)$,
we find that, for large
but finite $s$, the longitudinal propagator between the different
end-points of $V_+$ and $V_-$ becomes essentially independent of
the quark mass $m$ and the proper time cut-off $T$.
It takes the following form
\be
\label{minv}
\MiAB = 
\left(\begin{array}{cc} \log(se^{i\pi}) & \log s \\[1.5mm]
\log s & \log (se^{i\pi})
\end{array}\right)
\ee
This propagator is schematically depicted in fig 3.
In this way we find the following `regularized' form of $\MAB$
\ba
\label{mreg}
M_{reg}^{{}^{\! AB}} \is \left(\begin{array}{cc} 1+\eps & - 1+\eps \\
- 1 + \eps&  1 + \eps \end{array}\right) \\[2mm]
\eps^{-1} \is 1 - {2 i \over \pi}  \log s
\ea
We want to stress that this derivation of
the $\log s$-dependence is the direct translation to our
situation of the corresponding calculations done in perturbation theory.
Equation (\ref{mreg}), together with eqns (\ref{s12}),
(\ref{exp}) and (\ref{toedi}), completes the description
of our two-dimensional model for high-energy QCD.

\newsection{Comparison with standard perturbation theory}

It is an important and non-trivial test on our model to find out if it
is consistent with the known perturbative results \cite{lipetal,chengwu}.
In Appendix B we have summarized the known expression for the quark-quark
scattering amplitude to first order in $\log s$. The group factors that
receive a contribution to this order are depicted in fig 4.
We will now indicate
how these results can be derived from our effective two-dimensional action.
The results (\ref{answer}) have been obtained in the Lorentz gauge, so
we will choose a corresponding gauge  $\partial_i a_i^\pm=0$.
It is further useful to choose the following parametrization of the fields
\ba
\qquad g_2 g_1^{-1} \is \exp({e \theta}) \qquad \qquad \quad
g_1 h_1^{-1} \, =\, \exp(\tilde{e} \, \chi) \qquad \qquad
\nonu
h_2 h_1^{-1} \is \exp(e \phi) \qquad \qquad \qquad \ a_i^\pm \, =\,
{\tilde{e}\over 2}
\epsilon_i^j \partial_j \alpha^\pm
\ea
where
\be
\tilde{e}^2 = {2e^2\over \pi} \log s
\ee
The advantage of this parametrization is that vertex operators
that enter in the expression for the quark-quark amplitude
are simple exponentials
\be
\label{vert}
f(s,t) = {i s\over 2m^2} \int \! d^2 z \, e^{-iq\cdot z} \, \Bigl\langle
\, e^{ie\theta^a
\tau_{{}_{\! L}}^a}(0) \, \,
e^{ie\phi^b\tau_{{}_{\! R}}^b}(z) \, \Bigr\rangle
\ee
When we insert the above parametrization into the action $S$, expand into the
first two orders in $\tilde{e}$, we get
\ba
S \is {1\over 2}
\int d^2 z \, \tr\bigl(\partial_i \theta \partial_i \phi +
i (\partial_i \chi)^2 + i \partial_i \alpha^+\partial_i\alpha^-\bigr)
\nonumber \\[3mm]
& & + {\tilde{e}\over 2}
\int d^2 z \, \tr\bigl(\, \chi \, [\partial_i \theta,\partial_i \phi] \,
+ \, \half (\alpha^+\!+\!\alpha^-) \,
[\partial_i \theta, \partial_j \phi] \epsilon^{ij}\bigr)
\\[3mm] & & + {\tilde{e}^2 \over 4} \int d^2 z \, \tr \Bigl(
[\chi,\partial_i\theta] \, [\chi,\partial_i \phi] + \half [\partial_i
\alpha^+,\theta] \, [\partial_i \alpha^-, \phi] \Bigr) \nonumber
\ea
Here we dropped terms proportional to
$e$ since these are subdominant for large $s$.
{}From this form of the action it is clear that the scattering amplitude
indeed has a $\log s$ expansion of the form (\ref{sexp}).
The first terms in this perturbation expansion are obtained by evaluating
the diagrams of fig 4, and the one-loop corrections to the first two graphs.

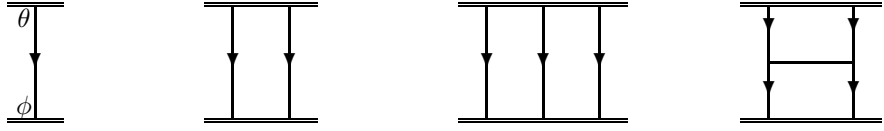
\fijn{Fig 4. These diagrams are the leading order diagrams
that contribute to the group factors $G_i$ and $F_2$ as given in
(\ref{group}).
The double horizontal lines represent the two quarks, and the thick
vertical lines are  $\theta$-$\phi$ propagators.}

As an example we outline the calculation of the
$H$-diagram of fig. 4, because this diagram illustrates in
a rather direct way the relation between our approach and the standard
theory of \cite{lipatov,chengwu}.
The standard result for this diagram is a sum of many contributions that
can be summarized in terms of Lipatov's effective gluon emission vertex
(see eqn (\ref{lipvert}) as \cite{lipatov}
\be
\label{AK}
A(s,q)  =
N {e^2\over (2\pi)^2} s \log s \int {d^2k_1 d^2k_2  K(k_1,k_2) \over
k_1^2\, k_2^2 \, (q\mint k_1)^2 (q\mint k_2)^2}
\ee
with
\ba
\label{Kkk}
K(k_1,k_2) \is C_\mu(k_1,k_2) C_\mu(q\mint k_1,q\mint k_2) \nonu
 \is -q^2 + {k_1^2(q\mint k_2)^2\over (k_1\mint k_2)^2} +
{k_2^2(q\mint k_1)^2 \over (k_1 \mint k_2)^2}
\ea
In (the above parametrization of) our two-dimensional effective field
theory, we only need to
add two Feynman diagrams to obtain the full answer for this amplitude,
namely the $H$-diagram with an intermediate $\chi$ and $\alpha =
\half (\alpha^+ +\alpha^-)$ line, resp. From the form (4.8) of the
action it is readily seen that the sum of these diagrams gives
\be
A(s,q) =  N { e^2 \over (2\pi)^2} s \log s \int {d^2k_1 d^2 k_2 \;\;
L(k_1,k_2)
\over k_1^2\,  k_2^2 \, (q \mint k_1)^2 (q\mint k_2)^2 (k_1\mint k_2)^2}
\ee
with
\be
L(k_1,k_2) = 2 [(k_1\cd k_2) (q\mint k_1)\cd(q\mint k_2) +
(k_1\cd \tilde{k}_2 )(q\mint k_1)\cd (\tilde{q}\mint \tilde{k}_2)]
\ee
with $\tilde{k}_{1,i} = \epsilon_{ij} k_{1,j}$ etc. This expression
is indeed identical to the standard result (\ref{AK}), because
\be
L(k_1,k_2) = {K(k_1,k_2) (k_1\mint k_2)^2}
\ee
This identity is most easily established by checking that both
sides coincide for all points where two momenta are equal or one
of them vanishes.

This correspondence with the standard theory
gives strong evidence that our approximations and reduction method
is correct and in principle can be extended to higher orders.
Because we have only done computations up to first non-trivial
order in $\log s$,  we have in particular not
yet seen any indication of the remarkable exponentiation
leading to the well-known reggeization of the gluon
propagator. It would be interesting to see if this result can be
derived from first principles, possibly by extending the methods
described here.

\newsection{Concluding Remarks}

\noindent
We have given a simple physical description of
the high energy interactions between two quarks, and shown that it
can be described in terms of a two-dimensional sigma-model action, given by
(\ref{toedi}). We have obtained this action by evaluating all correlations
in the longitudinal direction and keeping only the leading order terms
in $\lambda = t/s$. To obtain 1-loop corrections to this leading order
result one would need to evaluate the functional determinants (\ref{logdet}).
This will give rise to the usual renormalization of the coupling $e$.
The physical regime where one can expect this description to be relevant is
at short length and time scales, that is $s >\! > t > \Lambda_{qcd}$.
At these scales confinement has not yet set in, and the color electric
field behaves more like an ordinary electro-magnetic field. We should point
out, however, that our model fully incorporates the non-abelian character
of the gauge fields.

While most of the complicated dynamics of QCD is
eliminated in the limit $\lambda \ra 0$, our model may still contain
useful information about the strong coupling regime. In this respect,
it would be interesting
to further analyze the transversal dynamics described by (\ref{toedi}).
A lot is known about the more conventional
two-dimensional sigma-models, and most of this technology
can be taken over to our model, even though it has some
unusual features. The theory is perturbatively renormalizable and
presumably asymptotically free at short transverse distances.
Hence for relatively large values of $t$ perturbation theory
is expected to be reliable.

This simple model will of course not be able to describe
the full complexity of QCD, and has to be embedded into a more
elaborate and sophisticated framework before it can be used as a
realistic theory of high energy scattering of hadrons. One could
in particular imagine extending our model to include the scattering of
gluons and, instead of taking the trace of the Wilson lines,
folding the amplitude with a (phenomenological) hadron wave-function
that describes the distribution of the quarks and gluons in the
transverse space.
As a further comment we like to mention that the
rescaled 3$+$1 dimensional action (\ref{Sred}) in principle
can be used to compute gluon production in high-energy
collisions. Instead of integrating  out the transverse gauge field
$A_i$ to get to the two-dimensional model,
we can keep the gluons in our description by representing
the amplitude in the gluon Fock space. This presumably leads to a 2$+$1
dimensional model analogous to the extended eikonal model
(see e.g. \cite{chengwu}), where in addition to the transverse coordinates
$\xper$ one keeps the rapidity of the produced gluons as an extra coordinate.

\medskip

\noindent
{\bf Acknowledgements.}

\noindent
We would like to thank D. Amati, C. Callan, M. Ciafaloni,
D. Gross, L. Lipatov, D. Politzer, A. Polyakov and G. Veneziano for helpful
discussions. The research of H.V. is supported by NSF Grant PHY90-21984.
E.V. is supported in part by an Alfred P. Sloan Fellowship
and a Fellowship from the Dutch Royal Academy of Arts and Sciences.
\renewcommand{\thesection}{A}
\renewcommand{\thesubsection}{A.\arabic{subsection}}

\vspace{12mm}
\pagebreak[3]
\setcounter{section}{1}
\setcounter{equation}{0}
\setcounter{subsection}{0}
\setcounter{footnote}{0}

\begin{flushleft}
{\bf Appendix A: Non-abelian shock-waves.}
\end{flushleft}

\noindent
In this appendix we give
a short discussion of the classical shock wave field due to
a fast-moving non-abelian
charge. One way to obtain such solutions is to start
  with the classical field for a colored
particle at rest, and then apply an infinite Lorentz boost. The precise form
of the shock-wave will then
of course depend on the static configuration one starts with.
Whereas in the abelian theory there is only one candidate, the spherically
symmetric Coulomb field, it is known that the non-abelian Yang-Mills
equations also have other static solutions with less symmetries and
a smaller energy. The same ambiguity shows up here.

Consider the Yang-Mills equations in the
presence of a classical source
\be
\label{ymeqn}
D^\mu F_{\mu\nu}=j_\nu .
\ee
We are interested in the case where
the source is given by a classical point charge
that moves at the speed of light. Let the velocity of the particle be
in the $x^-$-direction and its transversal coordinate be $z\! = \! 0$,
then the only non-vanishing component of the source
$j_\nu$ is
\be
\label{fastp}
j_+=\lambda\delta(x^+)\delta^{(2)}(\xper)
\ee
Here  $\lambda$ is a constant element
of the lie algebra of the gauge group, describing the classical non-abelian
charge of the particle.

We can restrict the form of the solutions of the Yang-Mills equations
(\ref{ymeqn}) by using some of the symmetries of the source $j_\nu$.
In particular, we can require that the solutions we are looking for
must be invariant under translations in the $x^-$-direction and
boosts in the $(x^-,x^+)$ plane that leave the hyperplane $x^+\! =0$ fixed.
These two invariances are already very restrictive: they imply that all
fields are independent of $x^+$ and $x^-$, except for a possible
discontinuity at $x^+\! =0$.

We will restrict ourselves to the simplest type of solutions,
which are obtained by assuming that all components of
the Yang-Mills curvature vanish except $F_{+i}$.
These are then determined by
\ba
\label{ymm}
D^iF_{i+}\is \lambda\delta(x^+)\delta^{(2)}(\xper) \\[2mm]
D^+F_{i+} \is 0 \nonumber
\ea
To write this shock-wave we now choose the gauge $A_-=0$, and use the
residual gauge-invariance under $x^-$-independent gauge-transformations
to put also
$A_+=0$ everywhere. Further, since we assume that $F_{ij}=0$, the
transversal component $A_i$ must be pure gauge as well. Thus the
shock-wave can be represented as
\be
\label{shockw}
A_i =
\left\{\begin{array}{cc} 0 & \qquad \quad {\mbox{for}} \ \ \ x^+<0\\
g^{-1}\partial_ig & \qquad \quad {\mbox{for}} \ \ \  x^+>0 \end{array} \right.
\ee
where group element $g$ depends only on the transversal coordinate $\xper$.
Substituting this into (\ref{ymm}) gives
\be
\label{eom1}
\hat{D}_i(g^{-1}\partial_i g)=\lambda \delta^{(2)}(\xper).
\ee
Here, the covariant derivative involves the value of the gauge-field
$A_i$ at $x^+\!=\!0$, \ie at the position of the shock-wave
\be
\hat{D}_i = \partial_i + e[{A_i}, \ \ \ ]_{\strut{|x^+=0}}
\ee
So at this point the internal structure of the shock-wave becomes
relevant. In principle, the $A_i$ gauge field
at $x^+ = 0$ can be different from the field outside the shock wave.
So we can take
\be
\label{kalm}
A_i=h^{-1}\partial_i h \qquad \quad {\mbox{at}} \ \ \  x^+ = 0
\ee
and for any $A_i$ of this form
we can solve (\ref{eom1}) and obtain a classical
solution of (\ref{ymeqn}) for the source (\ref{fastp}).

When the gauge-field $A_i$ at $x^+\! =\! 0$
commutes with $g^{-1}\partial_ig$ in (\ref{eom1}) the shock-wave
takes the `abelian' form
\be
\label{gab}
g_{abelian}(z)=  \exp({e^2\over 4\pi} \lambda\log |\xper|^2)
\ee
This solution is uniquely singled out if we require that the classical field
respects all symmetries of the source.

The physical interpretation of the above field configuration becomes
more apparent after we perform a discontinuous gauge rotation and
put $A_i=0$ everywhere. After this $A_+$ will acquire a delta-function
singularity at the null-hyperplane $x^+=0$, such that the Wilson lines
that cross this plane are given by
\be
\label{step}
{\rm P}\exp(e \int_{\! -\epsilon}^{\epsilon} \!\! dx^+ \,
A_+) =g(\xper)
\ee
where $g(\xper)$ solves (\ref{eom1}). Physically this means that the
only physical effect of the shock wave is that when a charged
test particle passes through $x^+\! =0$ its wave function $\psi$ will
be instantaneously gauge-transformed to $\psi^\prime = g(\xper)\psi$.

\bigskip

\bigskip

\renewcommand{\thesection}{B}
\renewcommand{\thesubsection}{B.\arabic{subsection}}

\vspace{12mm}
\pagebreak[3]
\setcounter{section}{1}
\setcounter{equation}{0}
\setcounter{subsection}{0}
\setcounter{footnote}{0}

\begin{flushleft}
{\bf Appendix B: Perturbative Results to Order $e^6$.}
\end{flushleft}
In this appendix we give the known perturbative result of the high energy
quark-quark scattering amplitude, computed in \cite{lipetal,chengwu}.
It is known that the quark-quark
amplitude $f(s,t)$ can be expanded in a power series in
$\log s$
\be
\label{sexp}
f(s,t) =  s \sum_{n=0}^\infty A_n(q) ( \log s)^n
\ee
with $t= -q^2$.
Using standard perturbative techniques
the following results have been obtained for the first two
terms $A_0(q)$ and $A_1(q)$, upto order $e^6$ (we use the notation of
chapter 12 of
\cite{chengwu})
\ba
\label{answer}
A_0(q)  \is \ \
- {e^2 \over q^2} \; G_1 \ \ + \ \  {i e^4 \over 2!} I_2(q) \; G_2
\ \  + \ \   {e^6\over 3!} \, I_3(q) \; G_3  \qquad \\[4mm]
A_1(q) \is \, {N e^4\over 4\pi}  \, I_2(q) \; G_1
\, - \,  i{N e^6\over 4\pi} \, I_3(q)  \; G_2
\  + \ {e^6\over 4\pi} \,
\bigl[\, 2 I_3(q) - q^2 I_2^2(q)\, \bigr] \; F_2 \nonumber
\ea
where $G_k$ and $F_2$ are the group factors corresponding to the graphs
in fig 4.
\ba
\label{group}
G_k \is (\tau_{{}_{\! L}}^a \otimes \tau_{{}_{\! R}}^a)^k
\qquad \qquad \\[2mm]
F_2 \is (f_{ab c} \tau_{{}_{\! L}}^b \otimes \tau_{{}_{\! R}}^c)^2
\nonumber
\ea
where $\tau_{{}_{\! L}}^a$ and $\tau_{{}_{\! R}}^a$ are the non-abelian charges
of the left resp. right-moving quark,
and $I_2(q)$ and $I_3(q)$ are the (infrared divergent) integrals
\ba
I_2(q) \is \int
{d^2k\over(2\pi)^2} {1 \over k^2 (q-k)^2}
\\[3mm]
I_3(q) \is \int {d^2 k_1\over (2\pi)^2}{d^2 k_2\over (2\pi)^2}
{1 \over k_1^2 \, k_2^2 \, (q\! -\! k_1\! -\! k_2)^2} \nonumber
\ea
These result were obtained in the Lorentz gauge.


{\renewcommand{\Large}{\normalsize}
}

\end{document}